# Symbolic Methodology in Numeric Data Mining: Relational Techniques for Financial Applications

**Boris Kovalerchuk**
Central Washington University, Ellensburg, WA, USA, borisk@cwu.edu
**Evgenii Vityaev**
Institute of Mathematics, Russian Academy of Science, Novosibirsk, Russia, vityaev@math.nck.ru
**Husan Yusupov**
ABS, False Church, VA, USA, husan@earthlink.com

**Abstract**

*Currently statistical and artificial neural network methods dominate in financial data mining. Alternative **relational (symbolic) data mining** methods have shown their effectiveness in robotics, drug design and other applications. Traditionally symbolic methods prevail in the areas with significant **non-numeric (symbolic) knowledge**, such as relative location in robot navigation. At first glance, stock market forecast looks as a pure numeric area irrelevant to symbolic methods. One of our major goals is to show that **financial time series can benefit significantly from relational data mining based on symbolic methods**. The paper overviews relational data mining methodology and develops this techniques for financial data mining.*

## 1 Introduction

Historically, methods based on **attribute-value languages (AVLs)** have been most popular in applications of learning algorithms. One of the reasons is that in many areas including finance training data are naturally described by attributes of individual entities such as price, amount and so on. On the other hand relations between entities, such as "X is more expensive than Y", "A is a parent of B" and so on can be very useful for data mining. Sometimes the actual prices of X and Y are not available and relation Price(X) > Price (Y) is the only data available. Well-known, relatively simple and efficient **neural networks** and **decision trees** methods are typical examples of methods based on AVLs. However, these methods have serious **limitations in capturing relations**. Learning systems based on symbolic **first-order logic (FOL)** representations capture relations naturally. These methods have been successfully applied to many problems in chemistry, physics, medicine and other fields [Bratko et al., 1992; Muggleton et al., 1992; Muggleton, 1999; Bratko, 1993; Dzeroski et al., 1994; Kovalerchuk et al., 1997; Pazzani, 1997]. Dzeroski [1996], Bratko, Muggleton [1995], Muggleton [1999] and Pazzani [1997] listed some major successful applications of FOL. It was stated in these publications that the results obtained with relational methods using real industrial or environmental data are better than with any other known approach, with or without machine learning. Such tasks as mesh design, mutagenicity, and river water quality exemplifies successful applications. Domain specialists appreciate that the learned regularities are **understandable** directly in domain terms. Financial applications can specifically benefit from these methods. Fu [1999] noted "Lack of comprehension causes concern about the credibility of the result when neural networks are applied to **risky domains**, such as patient care and financial investment".

Traditionally, FOL methods were pure **deterministic techniques**, which originated in logic programming. There are well-known problems with deterministic methods--handling data with a significant level of **noise.** This is especially important for financial data, which typically have a very high level of noise. To utilize advantages of human-readable forecasting rules produced in relational data mining, **logical relations (predicates)** should be developed for financial problems. These predicates should be **interpretable** in ordinary financial terms like stock prices, interest rates, trading days, and so on. In this way, relational methods can produce valuable understandable rules in addition to the forecast.

Using this technique a financial specialist can **evaluate the performance of the forecast as well as a forecasting rule**. The problem of inventing predicates is addressed in this paper and in [Kovalerchuk, Vityaev, 2000, chapter 5]. Hybridizing the pure logical relational data mining methods with a probabilistic approach ("probabilistic laws") is a promising direction. This is done by introducing probabilities over logical formulas [Carnap, 1962; Fenstad, 1967; Vityaev E. 1983; Halpern, 1990, Vityaev, Moskvitin, 1993; Muggleton, 1994, Vityaev et al, 1995; Kovalerchuk, Vityaev, 1998, 2000, Koller, Pfeffer, 1997]. The MMDR method (section 6.4) is one of the few Hybrid Probabilistic Relational Data Mining methods developed and applied to financial data [Kovalerchuk, Vityaev, 1998, 2000; Vityaev et al, 1995; Vityaev, Moskvitin, 1993; Vityaev E., 1983] . The MMDR method has been applied to predict SP500C time series and to develop a trading strategy. This method outperformed several other strategies in simulated trading [Kovalerchuk, Vityaev, 1998, 2000].

Below we outline the FOL approach and describe a new **hybrid relational and probabilistic technique** that handles **numerical data** efficiently. In general the next generalization of relational data mining methods should handle **(1) classification, (2) interval and (3) numerical forecasting tasks with noise**. This is especially important in financial applications with numerical data and a high level of noise. We advocate **relational learning mechanisms**, which combine advantages of rule induction, analytical learning and statistical paradigms such as **statistical significance, explanatory power** and a **highly-expressive language**. Specifically rule induction and analytical methods have strong capabilities for explaining discovered patterns, statistical methods ensure reliability of these patterns and the analytical methods use a highly-expressive language (first-order predicate language) to ensure that complex patterns will not be overlooked.

The emphasis of our study is on development of numerical relational methods including relational representation of numeric data. How can one move from a real numerical measurement to a first-order logic representation? This is a non-trivial task [Krantz et al., 1971, 1989, 1990]. For example, how does one represent temperature measurement in terms of first-order logic without losing the essence of the attribute (temperature in this case) and without inputting unnecessary conventional properties? For instance, Fahrenheit and Celsius zeros of temperature are arbitrary conventions in contrast with the Kelvin scale where zero is the lowest possible temperature (the physical zero). Therefore incorporating properties of the Fahrenheit zero into first-order rules may force us to discover/learn properties of this convention along with more significant scale invariant forecasting rules. Learning algorithms in the space with those kind of arbitrary properties may be very time consuming and may produce inappropriate rules.

Table 1 summarizes the advantages and disadvantages of AVL-based methods and symbolic/relational first order logic (FOL) methods [Bratko, Muggleton, 1995]. Bratko and Muggleton [1995] pointed out that existing FOL systems are relatively inefficient and have rather **limited facilities for handling numerical data.** The purpose of new symbolic **Relational Data Mining (RDM)** is to overcome these limitations of current FOL methods.

*Table.1.* Comparison of AVL-based methods and first-order logic methods

| Method | Advantages for the learning process | Disadvantages for the learning process |
|---|---|---|
| Methods based on attribute-value languages | **Simple**, **efficient**, and handle **noisy data**. | **Limited form** of background **knowledge**. **Lack of relations** in the concept description language. |
| Methods based on First Order Logic | Appropriate learning time with a large **number of training examples**. Solid **theoretical basis** (first-order logic, logic programming). **Flexible** form of background knowledge, problem representation, and problem-specific constraints. **Understandable representation** of background knowledge, and relations between examples. | Inappropriate learning time with a large **number of arguments** in the relations. **Weak facilities** for processing **numerical data.** |

We use this new term, RDM, in parallel with the earlier terms **Inductive Logic Programming (ILP)** and First Order Logic (FOL) methods to emphasize the goal -- discovering **relations**. The terms ILP and FOL reflect the technique for discovering relations -- logic programming and FOL. In particular, discovering relational regularities can be done **without logical inference** and in languages of higher order. Therefore, we define **Relational Data Mining** as

*Discovering hidden relations (general logic relations) in numerical and symbolic data using background knowledge (domain theory).*

FOL systems have a mechanism to represent background financial knowledge in **human-readable and understandable form**. This is important for investors. Obviously, understandable rules have advantages over a stock market forecast without explanations. On the other hand, RDM should handle imperfect (noisy) data and in particular **imperfect numerical data.** One of the major obstacles to more effective use of the FOL methods is their limited facility for handling **numerical data** (see table 1). This is one of the active topics of modern RDM research [e.g., Bratko, Muggleton, 1995; Kovalerchuk, Vityaev, 2000, 1998; Vityaev et al., 1995].

There are two types of numerical data in data mining: (i) the numerical target variable and (ii) numerical attributes used to describe objects and discover patterns. Traditionally FOL methods solves only classification tasks without direct operations on numerical data. The MMDR method (section 6.4) handles an interval forecast of numeric variables with continuous values like prices along with solving classification tasks. In addition, MMDR handles **numerical time series** using the first-order logic technique, which is not typical for ILP and FOL applications.

**Statistical significance** is another challenge for deterministic methods. Statistically significant rules have an advantage in comparison with rules tested only for their performance on training and test data [Mitchell, 1997]. Training and testing data can be too limited and/or not representative. If rules rely only on them then there are more chances that these rules will not deliver a correct forecast on other data. This is a hard problem for any data mining method and especially for deterministic methods including deterministic ILP. We address this problem in Section 6. Intensive studies are being conducted for incorporating a probabilistic mechanism into ILP [Muggleton, 1994].

**Knowledge Representation** is an important and informal initial step in relational data mining. In attribute-based methods, the attribute form of data actually dictates the form of knowledge representation. Relational data mining has more options for knowledge representation. For example, attribute-based stock market information such as stock prices, indexes, and volume of trading should be transformed into the first order logic form. This knowledge includes much more than only values of attributes. There are many ways to represent knowledge in the first order logic language. One of them can skip **important information**; another one can hide it. Therefore, data mining algorithms may work too long to "dig" relevant information or even may produce inappropriate rules. Introducing **data types** [Flash et al., 1998] (see section 4) and concepts of **representative measurement theory** [Krantz et all, 1971, 1989, 1990; Pfanzagl, 1968] into the knowledge representation process helps to address this representation problem. In fact the measurement theory developed a wide set of data types, which cover data types used in [Flash et al., 1998].

It is well known that the general problem of rule generating and testing is **NP-complete** [Hyafil, Rivest, 1976]. Therefore, the discussion above is closely related to the following questions. What determines the number of rules? When do we stop generating rules? **The number of hypotheses** is another important parameter. It has already been mentioned that RDM with first order rules allows one to express naturally a large variety of general hypotheses, not only the relation between pairs of attributes. These more general rules can be used for classification problems as well as for an interval forecast of a continuous variable. RDM algorithms face exponential growth in the number of combinations of predicates to be tested. A mechanism to decrease this set of combinations is needed. Section 6 addresses these issues using a data type system and the representative measurement theory approach. Type systems and measurement theory approaches provide better ways to generate only **meaningful hypotheses** using syntactic information. A probabilistic approach also naturally addresses knowledge discovery in situations with **incomplete or incorrect domain** knowledge. In this way the properties of individual examples are not generalized beyond the limits of statistically significant rules.

## 2. Examples

Examples in this section illustrate (1) attribute-value object representation, (2) first order logic rules with one and two arguments in finance and (3) the difference between IF-Then first order logic rules and more traditional IF-Then propositional logic rules in finance.

**Example 1.**
Table 2 illustrates an attribute-value object representation. The first two lines represent some objects from a training data set. The last line represents an object without a value for the target attribute. The target value needs to be predicted for this object, i.e., stock price for the next day, 01.05.99. Each attribute-value pair can also be written as a name of an attribute and its value.

*Table .2.* Attribute-value object presentation

| Attribute: date | Attribute 1: Stock price on date t | Attribute 2: Volume (number of shares) traded on date t | Attribute 3: Target-- stock price on date t+1 |
|---|---|---|---|
| Value: 01.02.99 | Value: $ 60.6 | Value: 1,000,000 | $53.8 |
| Value: 01.02.99 | Value: $ 53.8 | Value: 700,000 | $54.6 |
| Value: 01.03.99 | Value: $ 54.6 | Value: 800,000 | $56.3 |
| Value: 01.04.99 | Value: $ 56.3 | Value: 840,000 | |

For instance, the following **rule 1** can be extracted from table 2:
   *IF stock price today is more than $60 AND trade volume today is greater than 900,000*
   *THEN tomorrow stock will go down.*
This rule can be written more formally:
   *IF StockPrice(t)>$60 AND StockTradeVolume(t)> 900,000*
   *THEN Greater(StockPrice(t+1), StockPrice(t))*
**Rule 2** is also true for table 2:
   *IF stock price today is greater than stock price yesterday AND*
   *trade volume today is greater than yesterday THEN tomorrow stock price will go up.*
**Rule 2** also can be written more formally:
   *IF Greater(StockPrice(t), StockPrice(t-1)) AND*
   *Greater(StockTradeVolume(t),StockTradeVolume(t-1)) THEN StockPrice(t+1)>StockPrice(t)*
Note, actually rule 2 is true for table 2 because table 2 does not have examples contradicting this rule. However, table 2 has only one example (t=01.03.99) confirming this rule. Obviously, table 2 is too small to derive reliable rules. Table 2 and presented rules are used just for illustrating that **attribute-value methods** were not designed to **discover rules 1 and 2** from table 2 directly**.** Both rules involve relations between **two objects** (records for two trading days t and (t+1)): *StockPrice(t+1)>StockPrice(t)* and *Greater(StockTradeVolume(t),StockTradeVolume(t-1))*.
Special **preprocessing** is needed to **create additional attributes**, such as
$$\text{StockUp(t)} = \begin{cases} 1, \text{StockPrice (t- 1)} < \text{StockPrice (t)} \\ 0, \text{StockPrice (t- 1)} \geq \text{StockPrice (t)} \end{cases}$$
There is a logical equivalency between attribute StockUp(t) and relation Greater(StockPrice(t), StockPrice(t-1)) used in rule 1:   *StockUp(t) ⇔Greater(StockPrice(t), StockPrice(t-1))*.
Similarly to be able to discover rule 2 with attribute-value methods we need an additional attribute:
*VolumeUp(t) ⇔ Greater(StockTradeVolume(t),StockTradeVolume(t-1))*
Let us try to add relations like *Greater(StockTradeVolume(t),StockTradeVolume(t-i))*

with 2, 3, ...i days ahead to the set of attributes. In this case, we need to generate many attributes such as Volume_iUp(t) similar to VolumeUp(t) used for one day ahead. In this way, a very small task can become huge. In logic terms attributes StockUp(t) and VolumeUp(t) are **monadic (unary) predicates** (Boolean **functions** with only one argument). In other words, languages of attribute-value methods are languages of functions of one variables (e.g., StockPrice(t)) and monadic predicates (e.g., StockUp(t)).

The first order language differs from a propositional logic language mainly by the **presence of variables**. Therefore, a language of monadic functions and predicates is a first order logic language, but a very restricted language. A language of monadic functions and predicates was not designed to represent relations that involve two, three or more objects. The domain (background) knowledge that can be used in the learning process of attribute-value methods is of a very restricted form. Moreover, other relations from a database cannot be used in the learning process if they are not incorporated into a single attribute-value table [Dzeroski, 1996].

**Example 2.**
There is a lot of **confusion** about the difference between logical attribute-value methods and relational methods. At first glance, they do the same thing -- produce "IF-Then" rules and use logical expressions. Dzeroski [1995] presented a fragment of a relational database for the potential customers of an enterprise to illustrate the difference. Table 3 presents a similar fragment of a relational database for corporate credit card holders. We wish to discover patterns in this table useful for distinguishing potential new cardholders from those who are not.

An attribute-value learning system may use Age, Sex and the Number of supervised associates from table 3. In this way, the following two patterns could be discovered with **monadic functions** Num_of_Supervised(Person) and Potential-Cardholder(Person):
Rule 1: *IF Num_of_Supervised(Person) ≥100  THEN Corporate_Cardholder(Person)*
Rule 2: *IF Sex(Person)=F AND Age(Person) ≥ 38  THEN Corporate_Cardholder(Person)*

*Table 3.* Database Relation "Potential-Corporate-Credit-Cardholder" (Attribute-value table)

| Person | Age | Sex | Number of supervised associates | Corporate cardholder |
|---|---|---|---|---|
| Diana Right | 39 | F | 10 | Yes |
| Carol Peterson | 49 | F | 1000 | Yes |
| Barbara Walker | 24 | F | 20 | No |
| Cindy Peck | 47 | F | 20 | Yes |
| Peter Cooper | 35 | M | 100 | Yes |
| Stephen Baker | 54 | M | 200 | Yes |

*Table .4.* Database relation "Colleague-of" (Attribute-value table)

| Person (CEO) | Colleague (CFO) |
|---|---|
| Peter Cooper | Diana Right |
| Stephen Baker | Cindy Peck |

Using a first order language with a two-argument predicate Colleague-of(person, colleague) the following pattern can be found:
Rule 3: *IF Colleague-Of(Person, Colleague) AND Corporate_Cardholder(Person)*
        *THEN Corporate_Cardholder(Colleague).*
The last rule is much more meaningful, than the first two formal rules. Rules 1 and 2 are discovered in an isolated file (table 3), but rule 3 is discovered using two files simultaneously. Table 3 represents a **single relation** in relational database terms. Table 4 represents another single relation. To find regularity involving **records from both tables** we need to use more expressive **first-order language**. Mathematically, first order languages generate such relations with two, three and more **variables.**

# 3. Relational data mining paradigm

As we discussed in the previous section, attribute-value languages are quite restrictive in inducing relations between different objects explicitly. Therefore, richer languages were proposed to express relations between objects and to operate with objects more complex than a single tuple of attributes. Lists, sets, graphs and composite types exemplify complex objects. These more expressive languages belong to the class of first order logic languages (see definitions in [Russell, Norvig, 1995; Dzeroski, 1996; Mitchell, 1977]). These languages support **variables, relations,** and **complex expressions**. FOL methods can discover regularities using several tables (relations) in a database, such as Tables 3 and 4, but the **propositional approach** requires creating a **single table**, called **a universal relation** (relation in the sense of relational databases) [Dzeroski, 1995].

Relational data mining methods should be able to solve numerical **and interval forecasting tasks** along with classification tasks such as presented above. This requires modifying the concept of positive and negative training examples $E^+$ and $E^-$ and modifying the concept of deriving (inferring) training examples from background knowledge and a predicate. A relational algorithm, called MMDR is able to solve numerical and interval forecasting tasks. This algorithm operates with a set of training examples E. Each example is amended with a target value like is done in Table 2, where attribute #3 is a target attribute--stock price for the next day. This is a numerical value. There is no need for MMDR to make this target discrete to get a classification task. Therefore, more generally, a **deterministic relational data mining (RDM)** mechanism is designed for forecasting tasks, including classification, interval and numerical forecasting. Similar to the definition for classification tasks, for general deterministic RDM **background knowledge B** is expressed as:

- a set of predicate definitions,
- training examples E expanded with target values T (nominal or numeric), and
- set of **hypotheses** $\{G_k\}$ expressed in terms of predicate definitions.

Using this background knowledge a RDM system will construct a set of predicate logic formulas $\{H_i\}$ such that: *the target forecast for all the examples in E can be **logically derived** from B and the appropriate $H_i$*.

**Example 3**. Let us consider Rule 2 discovered from Table 2.
*IF Greater(StockPrice(t), StockPrice(t-1)) AND*
*Greater(StockTradeVolume(t) StockTradeVolume(t-1)) THEN StockPrice(t+1)>StockPrice(t)*
This rule represents logical formula H, and table 2 represents training examples E. These two sources allow us to derive the following logically for date (t+1)=(01.04.99):

$$StockPrice(01.04.99) > 54.6 \qquad (1)$$

assuming that t=(01.03.99). This is consistent with actual StockPrice(01.04.99)=56.3 for date 01.03.99. Rule 1 from the same Example 1 in Section 4.1.2 represents logical formula $H_1$, but this rule is not applicable to t=(01.04.99). In addition, other rules can be discovered from Table 2. For instance,
*IF StockPrice(t)<$60 AND StockTradeVolume(t)< $90000 THEN Greater($60,StockPrice(t+1))*
This rule allows us to infer

$$StockPrice(01.04.99) < 60 \qquad (2)$$

Combining (1) and (2) we obtain

$$60 > StockPrice(01.04.99) > 54.6 \qquad (3)$$

With more data we can narrow the interval (54.6, 60) for t=(01.04.99). A similar logical inference mechanism can be applied for t=(01.04.99) to produce a forecast for (t+1)=(01.05.99). This example illustrates one of the ideas used in the hybrid MMDR method for numeric interval forecast.

In contrast with the deterministic approach, in **Hybrid Probabilistic Relational Data Mining** background knowledge B is expressed as:
- A set of predicate definitions,
- Training examples E expanded with target values (nominal or numeric), and

- A set of **probabilistic hypotheses** {G$_k$} expressed in terms of predicate definitions.

Using this background knowledge a system constructs a set of predicate logic formulas {H$_i$} such that:
*Any example in E is derived from B and the appropriate H$_i$ probabilistically, i.e., **statistically significantly**.* Applying this approach to (3) *60>StockPrice(01.04.99)>54.6,* we may conclude that although this inequality is true and is derived from table 2 it is not a statistically significant conclusion. It may be a property of a training sample, which is too small. Therefore, it is risky to rely on statistically insignificant forecasting rules to derive this inequality.

## 4.    Theory of RDM

### 4.1.    Data types in relational data mining

A **data type** (type for short) in modern object-oriented programming (OOP) languages is a rich data structure, <A,P,F>. It consists of **elements** A={a$_1$,a$_2$,...a$_n$}, **relations** between elements (**predicates**) P={P$_1$,P$_2$,...P$_m$} and meaningful **operations** with elements F={F$_1$,F$_2$,...,F$_k$}. Operations may include two, three or more elements, e.g., c = a # b, where # is an operation on elements a and b producing element c. This definition of data type formalizes the concept of a **single-level data type**. For instance, a single-level graph structure ("stock price" data type) can be created with nodes reflecting individual stock prices and edges reflecting relations between stock prices (<, =, >). These graph structures (values of the data type) can be produced for each trading day -- StPr(1), StPr(2),..., StPr(t) -- generating a time series of graph structures. A **multilevel data type** can be defined by considering each element a$_i$ from **A** as a composite data structure (data type) instead of as an atom. To introduce a multilevel stock price data type, stocks are grouped into categories such as high-tech, banking and so on. Then relations (<, =, >) between the average prices of these groups are defined. Traditional attribute-value languages operate with much simpler single-level data types. Implicitly, **each attribute** in attribute-value languages reflects a **type**, which can take a number of possible values. These values are elements of **A**. For instance, attribute "date" has 365 (366) elements from 01.01.99 to 12.31.99. There are several meaningful relations and operations with dates: **<, =, >,** and middle(a,b). For instance, the operation middle(a,b) produces the middle date c=01.05.99 for inputs a=01.03.99 and b=01.07.99. It is common in attribute-value languages that a data type such as a date is given as an **implicit data type** (see example 4 below). Usually in AVLs, relations **P** and operations **F** are not expressed explicitly. However, such data types can be embedded **explicitly** into **attribute-value languages.**

**Example 4**. Let us consider data type "trading weekdays", where a set of elements A consists of {Mon, Tue, Wed, Thu, Fri}. We may code these days as {1,2,3,4,5} and introduce a distance ρ(a,b)=|a-b| between them using these numeric codes. For instance, ρ(Mon,Tue)=ρ(1,2)=|1-2|=1 and
ρ(Fri,Mon)=ρ(5,1)=|5-1|=4. The last distance is natural if both Friday and Monday belong to **the same week**, but if Monday belongs to the next week it would be more reasonable to assign ρ(Fri,Mon)=1, because Monday is the next trading day after Friday. This is a property of **cyclical scales**. Different properties of cyclical scales are studied in representative measurement theory [Krantz, et al. 1971, 1979, 1980]. The "trading weekdays" data type is a **cyclical data type**. This distance has several properties which are unusual for distances. For instance, it is possible that ρ(a,b)≠ρ(b,a),

Let us assume that weekday *a* always precedes weekday *b*. Under this assumption ρ(Fri,Mon) means a distance between current Friday and Monday next week, but ρ(Mon,Fri) means a distance between Mon and Fri during the same week. In this example the requirement that a precedes b was not defined explicitly. In [Kovalerchuk, 1975, 1976] we studied cyclical scales and suggested numeric and binary coding schemes preserving this property for a variety of cyclical scales.

A new **strongly typed programming language Escher** was developed to meet this challenge [Flach et al, 1998]. The Escher language is an important tool, which allows users to incorporate a variety of explicit data types developed in representative measurement theory into the programming environment. On the other hand, RDM can be successfully implemented using common languages like Pascal and C ++ [Vityaev, Moskvitin, 1993; Vityaev et al, 1995].

## 4.2. Relational representation of examples

Relational representation of examples is the key to relational data mining. If examples are already given in this form, relational methods can be applied directly. For attribute-based examples, this is not the case. We need to express **attribute-based examples** and their **data types** in relational form. There are two major ways to express **attribute-based examples** using predicates: (1) generate predicates for each value and (2) use projection functions (see below). Table 5 presents an attribute-based data example for a stock.

*Table .5.* Attribute-based data example

| Stock price, $ | Volume, x1000 | Date | Weekday | Stock Event |
|---|---|---|---|---|
| 54.6 | 3067.54 | 01.04.99 | Monday | New product |

**Generating predicates for each value**. To express stock price $54.60 from Table 5 in predicate form, we may generate predicate P546(x), such that P546(x)=true if and only if the stock price is equal to $54.60. In this way, we would be forced to generate about 1000 predicates if prices are expressed from $1 to $100 with a $0.10 step. In this case, the ILP problem will be intractable. Moreover, the stock price data type has not yet been presented with the P546(x) predicate. Therefore, additional relations to express this data type should be introduced. For example, it can be a relation between predicates P546(x) and P478(x), expressing a property that stock price 54.6 is greater than 47.8. To avoid this problem and to constrain the hypothesis language for RDM, the **projection function** was introduced [Flach et al, 1998]. This concept is described below.

**Representation of background knowledge**. ILP systems use two sorts of background knowledge: objects and relations between those objects. For example, objects are named by constants a,b,c and relations are expressed using these names -- P(a,b)=true and P(c,b)=false. Use of constants is not very helpful because normally names do not carry properties of objects useful for faster data mining. In the approach suggested in [Flach et al, 1998], this is avoided. An object is "named" by the collection of all of its characteristics (**terms**).

For instance, term representation of stock information on 01.03.1999 can be written as follows:
*StockDate(w)=01.03.1999 & StockPrice(w)=$54.60 & StockVolume(w)=3,067,540 &*
*StockWeekday(w)=Mon & StockEvent(w)="new product"*.

Here *StockPrice* is a **projection function** which outputs stock prices (value of StockPrice attribute). Only naming of subterms is needed. This representation of objects (examples) is convenient for adding new information about an object (e.g., **data types**) and **localizing** information. For instance, subterm "StockEvent" permits one to localize such entities as reported profit, new products, competitor activity, and government activity.

In the example above the following **data types** are used:
- type weekday = {Mon, Tue, Wed, Thu, Fri},
- type price,
- type volume,
- type date,
- type event ={reported profit, new product, competitor's activity, government activity, ....},
- type stock = {price, volume, date, weekday, event}.

Type event brings a description of event related to the stock, e.g., published three month profit, new product, competitor's activity. This can be as a simple text file as a structured data type.

The representation of an example then becomes the **term**
*Stock (54.6, 3067.54, 01.04.99, Mon, new product)*.

Notice that when using projection functions in addition to predicates it is possible, without the use of variables, to represent relational information such as the equality of the values of two attributes. E.g., projection function StockEvent together with the equality relation (=) are equivalent to predicate SameEvent(w,x): *SameEvent(w,x) ⇔ StockEvent(x)=StockEvent(w)*.

Thus, the distinction between different propositional and first-order learning tasks depends in part on the representation formalism.

**Strongly typed languages**. FOL systems use types to provide labels attached to logical variables. However, these are not the data type systems found in modern programming languages. All available literals in the Prolog language will be considered for inclusion if a naive refinement operator is used for Prolog [Flash et al, 1998]. These authors developed a new strongly typed ILP language, **Escher**, which employs a complex data type system and restricts the set of hypotheses by **ruling out many useless hypotheses**. The MMDR method (Section 4.8) employs another way to incorporate data types into data mining by adding a data type structure (relational system) into the background knowledge. Such a relational system is based on representative measurement theory (Section 4.10).

**Complex data types and selector functions**. Each data type is associated with a relational system, which includes: (1) cardinality, (2) permissible operations with data type elements, and (3) permissible relations between data type elements.

In turn, each data type element may consist of its own subelements with their types. **Selector functions** [Flash et al, 1998] serve for extracting subterms from terms. Without selector functions, the internal structure of the type could not be accessed. Projection for selecting the i-th attribute requires the tuple type and a list of components (attributes) of the tuple. A list of components (attributes) requires the length of the list and the set of types of components.

**The number of hypotheses.** The most important feature of strongly typed languages is that they not only restrict possible values of variables, but also more importantly **constrain the hypothesis language.**

Table 6 summarizes information about data type features supported by different languages: ordinary attribute-based languages, attribute-based languages with types, first-order logic languages with types and ILP languages based on Prolog. This table is based on analysis from [Flach et al, 1998].Strongly typed languages for numerical data are especially important for financial applications with prevailing numeric data.

**Single Argument Constraints**. Consider an example, the term stock(A,B,C,D,E) has a type definition of stock(price, volume, date, weekday, event). Having this type definition, testing rules with arguments such as (25.7, 90000, 01.04.99, 67.3, new product) is avoided because 67.3 does not belong to weekday type. Thus, this typing information is a useful simple form of background knowledge. Algorithms FOCL (Section 6.3) and MMDR (Section 6.4) take advantage of typing information. On the other hand, the well-known FOIL algorithm (Section 6.2) does not use type constraints to eliminate literals from consideration.

*Table .6.* Data types supported by data mining languages

| Supported features of object representation | Attribute-based language | Attribute-based language with types | First-order language with types | ILP based on Prolog language |
|---|---|---|---|---|
| Formally expressed data type context | No | Yes | Yes | Yes |
| Attribute-value tuples | Yes | Yes | Yes | No |
| Explicitly induced relations between tuples | No | Yes | Yes | Yes |
| Data types of attributes expressed as in modern object-oriented programming languages | No | Yes | Yes | No |
| Mechanism to restrict the set of possible hypotheses using data types | No | Yes | Yes | No |
| Representing objects by terms using projection function | No | Yes | Yes | No |

Typing can be combined with **localized predicates** to **reduce the search** space. For instance, a localized relation *Greater_dates(A,B)* can be introduced to compare only dates with type information *Greater_dates(date,date)* instead of a universal relation *Greater(item, item)*. Similarly, a localized relation *Greater_$(A,B)*, type information *Greater_$(price, price)* can be introduced and applied for prices. This localized typing avoids the testing of some arguments (literals). For instance the localized predicate Greater_dates(A, B) should not be tested for literals of types such as *Greater_dates(stockprice, stockprice), Greater_dates(stockprice, date),* and *Greater_dates(date, stockprice).*

More generally, let $\{T_i\}$ be the types of already used variables $\{x_i\}$ in predicate P. Predicate P should be tested for different sequences of arguments. If the type $T_i$ of the already used i-th argument of P contradicts the type of an argument $y_i$ suggested for testing P, then the testing of the sequence which involves $y_i$ can be eliminated. This is a correct procedure only if a predicate is **completely localized**, i.e., only one type of argument is allowed for $y_i$. It is the case for the predicate *Greater_dates*, but it is not for the original predicate Greater defined for any items. This consideration shows that typing information **improves background knowledge** in two ways: (1) adding predicates and clauses about data types themselves and (2) refining and adding predicates and clauses about objects (examples). In such situations, typing can in the best case exponentially reduce the search space [Flach et al, 1998]**.** FOCL algorithms (Section 6.3) illustrates the benefit of typing. FOCL algorithm tested 3240 units and 242,982 tuples using typing as compared to 10,366 units and 820,030 tuples without typing. This task contained [Pazzani, Kibler, 1992]:
- learning a predicate with six variables of different types and
- 641 randomly selected training examples (233 positive and 408 negative training examples).

Typing is very useful for data mining tasks with limited training data, because it can improve the **accuracy of the hypothesis** produced without enlarging the data set. However, this **effect of typing is reduced** as **the number of examples increases** [Flash et al, 1998; Pazzani, Kibler, 1992].

**Existential variables.** The following hypothesis, which consists of two rules, illustrates existential variables:

*IF (there exists stock w such that StockEvent(x)=StockEvent(w)) AND (Some other statement) THEN StockPrice(x)>StockPrice(w)*

and

*IF ($\exists w, z$ StockEvent(x)=StockEvent(w)=StockEvent(z)) THEN StockPriceP(x)>StockPriceP(z).* The variables w and z are called **existential variables**. The **number of existential variables** like w and z provides one of the measurements of the **complexity of the learning task**. Usually the search for regularities with existential variables is a computational challenge.

### 4.3. First-order logic and rules

This section defines some basic concepts of first order logic. A **predicate** is defined as a binary function or a subset of a set $D=D_1 \times D_2 \times \ldots \times D_n$, where $D_1$ can be a set of stock prices at moment t=1 and $D_2$ can be stock price at moment t=2 and so on. Predicates can be defined **extensionally**, as a list of tuples for which the predicate is true, or **intensionally**, as a set of **(Horn) clauses** for computing whether the predicate is true. Let *stock(t)* be a stock price at t, and consider the predicate *UpDown(stock(t), stock(t+1), stock(t+2)),* which is true if stock goes up from date t to date t+1 and goes down from date t+1 to date t+2. This predicate is presented extensionally in Table 7.

*Table 7.* UpDown predicate

| Stock(t) | Stock(t+1) | Stock(t+2) | Updown( , , ) |
|---|---|---|---|
| $34 | $38 | $35 | True |
| $38 | $35 | $35.50 | False |
| $35.50 | $36 | $34 | True |
| $36 | $37 | $38 | False |

and **intensionally** using two other predicates Up and Down:
$Up(stock(t),stock(t+1)) \& Down(stock(t+1),stock(t+2)) \rightarrow UpDown(stock(t),stock(t+1),stock(t+2))$,
where $Up(stock(t),stock(t+1)) \Leftrightarrow Stock(t+1) \geq Stock(t)$, and
$Down(stock(t),stock(t+1)) \Leftrightarrow Stock(t) \geq Stock(t+1)$.
Predicates Up and Down are given extensionally in Table 8.

*Table 8.* Predicates Up and Down

| Stock(t) | Stock(t+1) | Up( , ) | Down( , ) |
|---|---|---|---|
| $34 | $38 | True | False |
| $38 | $35.50 | False | True |
| $35.50 | $36 | True | False |
| $36 | $37 | False | True |

**A literal** is a **predicate** A or its **negation (¬A).** The last one is called **a negative literal**. An unnegated predicate is called a **positive literal**. A **clause body** is a conjunction $A_1 \& A_2 \& ... \& A_t$ of literals $A_1, A_2, ..., A_t$. Often we will omit & operator and write $A_1 \& A_2 \& ... \& A_t$ as $A_1 A_2 ... A_t$. **A Horn clause** consists of two components: a **clause head** ($A_0$) and a **clause body** ($A_1 A_2 ... A_i ... A_t$). A clause head, $A_0$, is defined as a single predicate. A Horn clause is written in two equivalent forms: $A_0 \leftarrow A_1 A_2 ... A_i ... A_t$, or $A_1 A_2 ... A_i ... A_t \rightarrow A_0$, where each $A_i$ is a literal. The second form is traditional for mathematical logic and the first form is more common in applications.

   **A collection** of Horn clauses with the same head $A_0$ is called a **rule.** The collection can consist of a single Horn clause; therefore, a **single Horn clause** is also called a rule. Mathematically the term collection is equivalent to the OR operator ($\vee$), therefore the rule with two bodies $A_1 A_2 ... A_t$ and $B_1 B_2 ... B_t$ can be written as $A_0 \leftarrow (A_1 A_2 ... A_t \vee B_1 B_2 ... B_t)$. A **k-tuple,** a **functional expression,** and a **term** are the next concepts used in relational approach. A finite sequence of k constants, denoted by $<a_1,...,a_k>$ is called a **k-tuple** of constants. A function applied to k-tuples is called a **functional expression**. A **term** is a constant, variable or (3) functional expression. Examples of terms are given in table 9. A k-tuple of terms can be constructed as a sequence of k terms. These concepts are used to define the concept of atom. An **atom** is a predicate symbol applied to a k-tuple of terms. For example, a predicate symbol P can be applied to 2-tuple of terms *(v,w),* producing an atom *P(v,w)* of arity 2. If P is predicate ">" (greater), *v=StockPrice(x)* and *w=StockPrice(y)* are two terms then they produce an atom: *StockPrice(x) >StockPrice(y),* that is, price of stock x is greater than price of stock y. Predicate P uses two terms v and w as its arguments. The number two is the **arity** of this predicate.

*Table 9.* Examples of terms

| Expression | Comment | Term ? |
|---|---|---|
| x | Variable --stock x | Yes |
| MSFT | Constant (specific stock/index) | Yes |
| StockPrice(x) | Functional expression | Yes |
| TradeVolume(x) | Functional expression | Yes |
| StockPrice(x)*TradeVolume(x) | Functional expression | Yes |
| Nasdaq(x)>StockPrice(x) | Incorrect | No |
| NASDAQ(x) | Predicate, literal (Stock x is traded on NASDAQ) | No |
| StockPrice(x)>StockPrice(y) | Predicate(x,y), literal | No |

If a predicate or function has k arguments, the number k is called **arity** of the predicate or function symbol. By convention, **function and predicate symbols** are denoted by Name/Arity. Functions may have **variety of values**, but predicates may have only Boolean values **true and false**. The meaning of the rule for a k-arity predicate is the set of k-tuples that satisfy the predicate. A tuple satisfies a rule if it satisfies one of the Horn clauses that define the rule. **A unary (monadic) predicate** is a predicate with arity 1. For example, NASDAQ(x) is unary predicate. Predicates defined by a collection of examples are called **extensionally defined predicates**, and **predicates** defined by a rule are called **intensionally defined predicates.** If predicates defined by rules then inference based on these predicates can be explained in terms of these rules. Similarly, the **extensionally defined predicates** correspond to the **observable facts** (or the **operational predicates**) [Mitchell, Keller, & Kedar-Cabelli, 1986]. A collection of intensionally defined predicates is also called **domain knowledge** or **domain theory.** Statements about a particular stock MSFT for a particular trading day can be written as:
*StockPrice(MSFT)>83, NASDAQ(MSFT), TradeVolume(MSFT)=24,229,000.*

## 5. Background knowledge

### 5.1. Arguments constraints and skipping useless hypotheses

Background knowledge fulfills a variety of functions in the data mining process. One of the most important is reducing the number of hypotheses to be tested to speed up learning and make this process tractable. There are several approaches to reduce the size of the hypothesis space. Below two of them are presented. They use constraints on arguments of predicates from background knowledge B. The difference is that the first approach uses **constraints on a single argument** and the second one uses **constraints on several arguments** of a predicate defined in B. The first approach called **typing** approach is based on information about individual data types of arguments of a predicate. For instance, suppose only an integer can be the first argument of predicate P and only the date (M/D/Y) can be the second argument of this predicate. It would be wasteful to test hypotheses with the following typing P(date, integer), P(integer, integer) and P(date, integer). The only one correct type here is P(integer, date). The second approach is called **inter-argument constraints approach.** For example, predicate *Equal(x,x)* is always true if both arguments are the same. Similarly, it is possible that for some predicate P for all x P(x,x)=0. Therefore, testing hypotheses extended by adding Equal(x,x) or P(x,x) should be avoided and the **size of the hypothesis space explored can be reduced**. The value of inter-argument constraints is illustrated by the experimental fact that the FOCL algorithm, using **typing and inter-argument constraints**, was able to test 2.3 times less literals and examples than using only typing. [Pazzani, Kibler, 1992]. Table 10 summarizes properties of the two discussed approaches for reducing the number of hypotheses.

*Table 10.* Approaches for reducing hypothesis space

|  | Approach 1: Implementing a single argument constraint (typing ) | Approach 2: Implementing inter-argument constraints |
|---|---|---|
| **Definition** | Properties of an individual argument of the predicate. | A relationship between different arguments of a predicate |
| **Example of constraints** | Only an integer can be the first argument of a predicate. Only date (M/D/Y) can be the second argument of the predicate. | All of the variables in one predicate should be different, i.e., a hypothesis should not include predicate P(x,x), but may include P(x,y) |
| **Experiment** | FOCL algorithm, using typing and inter-argument constraints, was able to test two times less literals and examples than using only typing [Pazzani, Kibler, 1992]. | |

### 5.2. Initial rules and improving search of hypotheses

This section considers another useful sort of background knowledge--a (possibly incorrect) partial initial rule that approximates the concept (rule) to be learned. There are two basic forms of this initial rule: (1) extensional form and (2) intensional form. If a **predicate is defined by other predicates**, we say the definition is **intensional.** Otherwise, a predicate given by example is called **extensional**. It is also possible that background knowledge B contains a predicate in a **mixed way** partially by examples and partially by other predicates. In general, background knowledge presented in a mixed way reduces the search. [Passani, Kibler, 1992].

   **Learning using initial extensional rule**. An expert or another learning system can provide an initial extensional rule [Widmer, 1990]. Then this rule (initial concept) is refined by adding clauses [Passani, Kibler, 1992]:

1. An algorithm computes the criterion of optimality (usually information gain) of each clause in **the initial concept**.
2. The literal (or conjunction of literals) with the maximum gain is added to the end of the current clause (start clause can be null).
3. If the current clause covers some negative tuples (examples), additional literals are added to rule out the negative tuples.

     **Learning using initial intensional rules**. Next, consider domain knowledge defined in terms of extensional and intensional initial predicates. Systems such as CIGOL [Muggleton & Buntine, 1988] make use of (or **invent**) **background knowledge** of this form. For example, if an extensional definition of the predicate *GrowingStock(x,y,z)* is not given, it could be defined in terms of the intensional predicate GreaterPrice by:

     *GrowingStock(x,y,z) ← GreaterPrice(x,y), GreaterPrice(y,z),*
where x, y, and z are prices of the stock for days t, t+1, and t+2, respectively.

     It is possible that the intensional predicate GrowingStock(x,y,z) added to the hypothesis improves it, but each of predicates GreaterPrice(x,y) and GreaterPrice(y,z) does not improve the hypothesis. herefore, common search methods may not discover a valuable stock regularity. Pazzani and Kibler [1992] suggested that if the literal with the maximum of the optimality criterion (gain) is intensional, then the literal is made extensional and the extensional definition is added to the clause under construction. Note that computation of the optimality criterion, which guides the search, is different for extensional and intensional predicates. For intensional predicates it is usually involves a Prolog proof. Potentially operationalization can generate very long rules.

     **Learning using initial intensional and extensional rules.** The previous consideration has shown that adding background knowledge can increase the ability of algorithms to find solutions. Table 11 shows an example of partial background knowledge for a stock market forecast. It consists of
- a definition of the target predicate UP(x,y,w,z) with four arguments to be learned,
- typing information about x,y,w and z,
- intensional predicate Q(x,y,w) with three arguments to be used for discovering predicate Up(x,y,w,z), and
- extensional predicates Monday(t) and Tuesday(t) to be used for discovering Up(x,y,w,z).

*Table 11.* Partial background knowledge for stock market

| |
|---|
| **Definition of target predicate to be learned:** <br> Up(Stock(t), Stock(t+1),Stock(t+2)). <br> IF Stock(t+2)) < Stock(t+3) THEN this predicate should be true and the predicate is false If Stock(t+2)) ≥ Stock(t+3). <br> Up(Stock(t), Stock(t+1),Stock(t+2))⇔Stock(t+2) < Stock(t+3) <br><br> To compute this predicate only stock prices Stock(t), Stock(t+1) and Stock(t+2)) can be used. Actually the predicate should forecast stock price for date t+3, having stock prices for the three preceding days. The learning algorithm should learn the predicate Up, .i.e., generate a logical rule combining Stock(t), Stock(t+1),Stock(t+2) such that <br> Up(Stock(t), Stock(t+1),Stock(t+2))⇔Stock(t+2) < Stock(t+3)  for all training data. <br><br> **Type :** UP(float, float, float, float) <br> Positive examples,  Pos: Ex1--(34.0, 35.1, 36.2, 37.4), Ex2--(37, 38.1, 34.4, 35.7) <br> Negative examples, Neg: Ex3--(33.2, 32.1, 33.7, 31.6), Ex4--(30.8 29.3, 28.8 27.9) <br><br> **Intensional Predicate(s):** <br> Q(Stock(t),Stock(t+1), Stock(t+2)) ⇔  Stock(t+1)-Stock(t) <Stock(t+2)-Stock(t+1) <br> Type : Q(float, float, float); <br><br> **Extensional Predicates:** <br> Monday (t).  t type: date. This predicate is true for Mondays. <br> Pos : (04.05.99)(04.12.99)(04.19.99)...(11.01.99) <br> Tuesday(t). t type: date.  This predicate is true for Tuesdays. <br> Pos : (04.06.99)(04.13.99)(04.20.99)...(11.02.99) |

In addition, Table 12 provides an initial intensional rule for the target concept Up(x,y,w,z). This rule assumes that if growth was accelerated from date t to t+2 then the stock will grow further on date t+3. Background knowledge is called **extended background knowledge** if it includes: (1) extensional knowledge (training examples and extensional predicates), (2) initial rules, and (3) intensional target concept definition.

*Table 12.* Intensional initial rule for the target concept

| Up(Stock(t); Stock(t+1);Stock(t+2) ← Q(Stock(t+1),Stock(t), Stock(t+2),Stock(t+1)) |
|---|

Pazzani and Kibler [1992] found in experiments that **extended background knowledge** with a **correct intensional target** definition avoids exhaustive **testing** every variable of every predicate and increases the speed of the **search**. In their experiments, a correct extensional definition of the target concept was found by testing only 2.35% of literals needed for rule discovery if the target concept is not provided. However, the same research has shown that extended background knowledge (1) can increase the **search space**, (2) can decrease the **accuracy** of the resulting hypothesis, if the background knowledge is partially irrelevant to the task, and (3) can increase the number of **training examples** required to achieve a given accuracy.

These observations show the need for **balancing initial intensional and extensional predicates** in background knowledge. One of them can be more accurate and can speed up the search for regularity more than other. Therefore, the following procedure will be more efficient:
- Compare accuracy of intensional and extensional knowledge.
- Include a more accurate one in the background knowledge.
- Discover regularities using the most accurate background knowledge from 2).
- Discover regularities using all background knowledge.

The modification of this mechanism includes use of **probabilities** assigned to all types of background knowledge (see Section 6) There are several ways to combine extensional and intensional knowledge in discovering regularities. One of them is converting initial rules (predicates) into extensional form (**operationalize a clause**) if it has positive information gain. The extensional predicates are compared to the induced literal with the maximum information gain. This approach is used in the FOIL algorithm (see Section 6.2). In an **explanation-based learning approach,** the target concept is assumed a **correct**, **intensional definition** of the concept to be learned and the domain knowledge is assumed correct as well. An approach that is more realistic is implemented in algorithms such as FOCL and MMDR. These methods relax the assumption that the target concept and the domain knowledge are correct.

## 6.  Algorithms: FOIL and FOCL

### 6.1.  Overview

A variety of relational machine learning systems have been developed in recent years [Mitchell, 1997]. Theoretically, these systems have many advantages. In practice though, the complexity of the language must be severely restricted, reducing their applicability. For example, some systems require that the concept definition be expressed in terms of **attribute-value pairs** [Lebowitz, 1986; Danyluk, 1989] or only in terms of **unary predicates** [Hirsh, 1989; Mooney, Ourston, 1989; Shavlik, Towell, 1989; Pazzani, 1989]. The systems that allow actual **relational concept definitions** (e.g., OCCAM [Pazzani, 1990], IOE [Flann & Dietterich, 1989], ML-SMART [Bergadano et al., 1989]) place strong restrictions on the form of induction and the initial knowledge that is provided to the system [Pazzani, Kibler, 1992].

In this section, we present three relational data mining methods: FOIL, FOCL and MMDR. Algorithm FOIL [Quinlan, 1990] learns constant-free Horn clauses, a useful subset of first-order predicate calculus. Later FOIL was extended to use a variety of types of background knowledge to increase the class of problems that can be solved, to decrease the hypothesis space explored, and to increase the accuracy of learned rules.

Algorithm FOCL [Pazzani, Kibler, 1992], already mentioned several times, extends FOIL. FOCL uses first order logic and FOIL's information-based optimality metric in combination with background knowledge. This is reflected in its full name -- First Order Combined Learner. FOCL has been tested on a variety of problems [Pazzani, 1997] that includes a domain theory describing when a **student loan** is required to be repaid [Pazzani & Brunk, 1990]. It is well known that the general problem of rule generating and testing is NP-complete [Hyafil, Rivest, 1976]. Therefore, we face the problem of designing NP-complete algorithms. There are several related questions. What determines the number of rules to be tested? When should one stop generating rules? What is the justification for specifying particular expressions instead of any other expressions? FOCL, FOIL and MMDR use different stop criteria and different mechanisms to generate rules for testing. MMRD selects rules, which are simplest and consistent with measurement scales [Krantz et all, 1971, 1989, 1990] for a particular task. The algorithm stops generating new rules when the rules become too complex (i.e., statistically insignificant for the data) in spite of the possibly high accuracy of the rules when applied to training data. The obvious other stop criterion is time limitation. FOIL and FOCL are based on the information gain criterion.

## 6.2. FOIL

The description of FOIL and FOCL below is based on [Pazzani, Kibler 1992]. FOIL uses positive and negative examples $\{e^+\}$, $\{e^-\}$ for some concept C, and related (background) predicates. FOIL tries to generate a rule R combining these predicates in such a way that R is true for positive examples, $R(e^+)=1$, and false for negative examples, $R(e^-)=0$. This rule should not contain constant and function symbols, but can contain **negated predicates** in both FOIL and FOCL. One of the FOIL design is shown in Table 13 [Pazzani, Kibler, 1992]. FOIL has two major stages: (1) the **separate stage,** which begins a new clause, and (2) the **conquer stage,** which constructs a conjunction of literals to serve as the body of the clause.

*Table 13.* FOIL Design 1

| |
|---|
| Let POS be the positive examples. |
| Let NEG be the negative examples. |
| Set NewClauseBody to empty. |
| Until POS is empty do: |
|     Separate: (**begins new clauses**) |
|         Remove from POS all examples that satisfy the NewClauseBody. |
|         Reset NEG to the original negative examples. |
|         Reset NewClauseBody to empty. |
|     Until NEG is empty do: |
|         Conquer: (**build clause body**) |
|             Choose a literal L. |
|             Conjoin L to NewClauseBody. |
|             Remove from NEG examples that do not satisfy L. |

Each clause describes some subset of the positive examples and no negative examples. FOIL uses two **operators**: (1) Start a new, empty clause, and (2) Add a literal to the end of the current clause. Adding literals continues until no negative example is covered by the clause. These literals are added to the end of the current clause. FOIL starts new clauses until all positive examples are covered by some clause. Adding literals assumes a mechanism to **generate literals**, i.e., a particular combination of variables and predicate names. If a predicate (predicate name) is already selected the choice of variables is called a **variablization** (of the predicate) [Pazzani, Kibler, 1992]. If the variable chosen already occurs in an unnegated literal of the rule, then the variable is called **old.** Otherwise, the variable is called **new**. FOIL and FOCL require at least one old variable. This old variable can be in either the head or the current body of the rule (Horn clause). FOIL uses hill climbing optimization approach to add the literal with the maximum information gain to a clause (rule). This requires computing the information gain for each variablization of each predicate P. The **information gain metric** used by FOIL is

$$\text{Gain(Literal)} = T^{++} * (\log_2 (P_1/P_1+N_1) - \log_2 (P_0/P_0+N_0)),$$

where $P_0$ and $N_0$ are the numbers of positive and negative tuples **before** adding the literal to the clause, $P_1$ and $N_1$ are the numbers of positive and negative tuples **after** adding the literal to the clause, and $T^{++}$ is the number of positive tuples before adding the literal that has at least one corresponding extension in the positive tuples after adding the literal [Quinlan, 1990].

**Cost.** A hill-climbing search used by FOIL depends on the branching factor of the search tree. This branching factor is an exponential function: (1) of the arity of the available predicates, (2) of the arity of the predicate to be learned, and (3) of the length of the clause that is being learned [Pazzani, Kibler, 1992]. Two measures estimate the cost of FOIL computation:
- the theory-cost--**the number of different literals** that can be chosen to extend the body of the given clause (does not depends on the number of training examples),
- evaluation-cost --the **cost of computing the information gain** of each literal (depends on the number of training examples).

**Heuristic**. Testing the variablizations of some predicates is avoided by a branch-and-bound pruning heuristic in FOIL. The idea is that some variablization can be more specific than another. In a more general variablization, an old variable is replaced with a new variable. The heuristic prefers a more general variablization, computing maximum possible information gain of a predicate with this variablization. An additional stopping criterion allows FOIL to learn from **noisy data.**

### 6.3. FOCL

Algorithm FOCL [Pazzani, 1997, Pazzani, Kibler, 1992] extends and modifies FOIL to permit the various forms of **background knowledge**: (a) **constraints** to limit the search space, (b) **predicates defined by a rule** in addition to predicates defined by a collection of examples, and (3) **input a partial, possibly incorrect rule** that is an initial approximation of the predicate to be learned.

These extensions guide construction of a clause by selecting literals to test. FOCL attempts to constrain search by using variable **typing**, **inter-argument constraints**, and an **iterative-widening approach** to add new variables. FOCL specification is presented in Table 14.

*Table 14*. FOCL Specification [Pazzani, Kibler, 1992].

| |
|---|
| Given:<br>    1. The name of a predicate of known arity.<br>    2. A set of positive tuples.<br>    3. A set of negative tuples.<br>    4. A set of extensionally defined predicates.<br>    5. A set of intensionally defined predicates (optional).<br>    6. A set of constraints (e.g., typing) on the intensional and extensional predicates (optional).<br>    7. An initial (operational or non-operational) rule (optional).<br>Create: A rule in terms of the extensional predicates such that no clause covers any negative examples and some clause covers every positive example |

**Summary of FOCL.** Authors of FOCL draw a number of important conclusions about the complexity of learning rules and the value of different sorts of knowledge. Some of these conclusions are summarized below:
- The branching factor grows exponentially in the arity of the available predicates and the predicate to be learned.

- The branching factor grows exponentially in the number of new variables introduced.
- The difficulty in learning a rule is linearly proportional to the number of clauses in the rule.
- Knowledge about data types provides an exponential decrease for a search necessary to find a rule.
- Any method (argument constraints, semantic constraints, typing, symmetry, etc.) that eliminates fruitless paths decreases the search cost and increases the accuracy.
- The uniform evaluation function allows FOCL to tolerate domain theories that are both incorrect and incomplete.

Advantages of FOCL were experimentally confirmed by Pazzani and Kibler (see section 5.1).

### 6.4. Algorithm MMDR

A **Machine Method for Discovering Regularities (MMDR)** contains several extensions over other RDM algorithms. It permits various forms of **background knowledge** to be exploited. The goal of the MMDR algorithm is to create probabilistic rules in terms of the relations (predicates and literals**)** defined by a collection of examples and other forms of background knowledge. MMDR as well as FOCL has several advantages over FOIL:

- Limits the search space by using **constraints.**
- Improves the search of hypotheses by using background knowledge with **predicates defined by a rule directly** in addition to predicates defined by a collection of examples.
- Improves the search of hypotheses by accepting as input **a partial, possibly incorrect rule** that is an initial approximation of the predicate to be learned.

There are also advantages of MMRD over FOCL:

- Limits the search space by using the **statistical significance** of hypotheses.
- Limits the search space by using the strength of **data types scales.**
- Shortens the final discovered rule by using the initial set of hypotheses in intensional form directly (without operationalization).

The advantages above represent a way of generalization used in MMDR. Generalization is the critical issue in applying data-driven forecasting systems. The MMDR method generalizes data through "**lawlike" logical probabilistic rules** presented in first order logic. .

Theoretical advantages of MMDR generalization are presented in [Vityaev, 1976, 1983, 1992, Vityaev, Moskvitin, 1993, Vityaev et al, 1995, Kovalerchuk, 1973, Zagoruiko, 1976, Samokhvalov, 1973]. This approach has some similarity with the hint approach [Abu-Mostafa, 1990]. The main source for hints in first-order logic rules is representative measurement theory [Krantz et al., 1971, 1989, and 1990]. Note that a class of general propositional and first-order logic rules, covered by MMDR is wider than a class of decision trees.

MMDR selects rules, which are simplest and consistent with measurement scales (data types) for a particular task. Initial rule/hypotheses generation for further selection is problem-dependent. In [Kovalerchuk, Vityaev, 2000, chapter 5], we presented a set of rules/hypotheses specifically generated as an initial set of hypotheses for financial time series. This set of hypotheses can serve as a catalogue of initial rules/hypotheses to be tested (learned) for stock market forecasts. The steps of MMDR are described in Figure 1. The first step selects and/or generates a class of logical rules suitable for a particular task. The next step learns the particular first-order logic rules using available training data. Then the first-order logic rules on training data using Fisher statistical test [Kendall, Stuart, 1977; Cramer, 1998] are tested. After that statistically significant rules are selected and Occam's razor principle is applied: the simplest hypothesis (rule) that fits the data is preferred [Mitchell, 1997, p. 65]. The last step creates interval and threshold forecasts using selected logical rules: *IF A(x,y,…,z) THEN B(x,y,…,z)*.

To address the NP-complete problem of rule generating and testing the MMDR method stops generating new rules, when rules become too complex. The rule is too complex is it is statistically insignificant for the data in spite of possible high accuracy of these rules for the training data. In this way, the problem becomes tractable. The obvious other MMDR's stop criterion is a time limitation. The original challenge for MMDR was the simulation of discovering **scientific laws** from empirical data in chemistry and physics. There is a well-know difference between "black box" models and fundamental models (laws) in modern physics. The latter have much longer life, wider scope, and a solid background. There is a reason to believe that MMDR caught some important features of discovering these regularities ("laws").

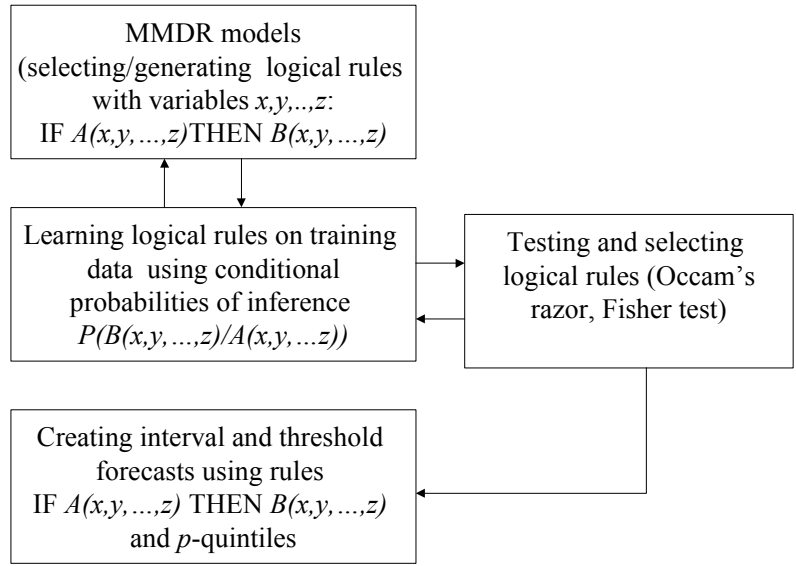

*Figure 1.* Flow diagram for MMDR: steps and technique

## 7. Experimental results

This section presents some examples of use of MMDR in prediction of SP500. Comparison of forecasting performance obtained by use of different methods is presented in three tables below. These data shows that MMDR outperformed other methods.

*Table 14.* ARIMA and MMRD

| Model # | Forecasting performance (correct buy/sell signal) | Comment |
|---|---|---|
| 1 | No forecast | Buy/sell strategy is based on non-zero difference between $T(t+1)$ and $T(t)$. This random walk model has zero difference for all days. Only random advice is possible here with 50% of success. |
| 2 | 62.58% | This model was selected without any connections with MMDR. Its forecast is less precise than produced by MMDR ( 0.7 and 0.84, respectively, correct up-down and down-up forecasts). |
| 3 | 58.84% | This model was selected without any connections with MMDR. It is less precise than the rules produced by MMDR |
| 4 | 79.6% | This simple Markov process was identified by MMDR search approach. All weekdays $t$ (Mon., Tue.,Wed., Thu. and Fri.) are tested for discovering values of parameters $s$ and $t$. |
| 5 | 75.92% | This model exploits parameters prompted by rules discovered by MMDR. Performance 75.92% is fully consistent with MMDR performance (70% and 84%, respectively, correct up-down and down-up forecasts). |

The most significant advantage of the first order methods and MMDR, in particular, is that they can **forecast directly the sign of the difference in SP500 instead of the value** as ARIMA does. ARIMA can generate a sign forecast using a predicted value. The forecast of a value is more complex and available data may not fit for value forecast. The value forecast can be inaccurate and statistically insignificant, but the forecast of the sign can be accurate and statistically significant for the same data.

*Table 15.* Forecast performance of different methods on test data

| Method | % of correct sign (up/down) forecast of SP500C | | |
|---|---|---|---|
| | 1995-1996 | 1997-1998* | Average 1995-1998 |
| Risk Free (3%) | N/A | N/A | |
| Neural network 1 (with preprocessing) | 68% | 57 | 62.5% |
| Rules extracted from NN 1(indirect estimate) | $\leq 68\%$ | $\leq 57\%$ | $\leq 62.5\%$ |
| Decision tree (Sipina with C4.5 simplification) | 67% | 60% | 64% |
| First-0rder logic with probability (MMDR) | 78% | 85% | 81.5% |
| First-order logic method (FOIL) | 50.50% | 45.40% | 47.95% |

\* Data for 1998 are used from 01.01.98 to 10.31.98.

*Table 16.* Simulated gain per year for SP500

| Method | Gain per year in simulated trading (% of initial investment) | | |
|---|---|---|---|
| | 1995-1996 | 1997-1998 | Average 1995-1998 |
| Adaptive Linear | 21.9 | 18.28 | 20.09 |
| MMDR | 26.69 | 43.83 | 35.26 |
| Buy-and-Hold | 30.39 | 20.56 | 25.47 |
| Risk-Free | 3.05 | 3.05 | 3.05 |
| Neural Network | 18.94 | 16.07 | 17.5 |

The most interesting is comparison of the MMDR with the Buy-and-Hold (B&H) strategy. B&H strategy slightly outperformed MMDR for 1995-1996 (30.39% for B&H and 26.69% for MMDR, see Table 16. On the other hand MMDR significantly outperformed Buy-and-Hold for 1997-1998 (43.83% for MMDR and 20.56% for B&H.)